\documentclass[%
twocolumn,
english,
 amsmath,amssymb,
prx,
longbibliography,
floatfix,
]{revtex4-1}

\usepackage{graphicx}
\usepackage{xcolor}
\usepackage{dcolumn}
\usepackage{bm}
\usepackage{dsfont}
\usepackage{appendix}
\usepackage[colorlinks=true,urlcolor=blue,citecolor=blue,linkcolor=blue]{hyperref} 
\usepackage{placeins}

\begin{document}

\title{Investigation of the N\'eel phase of the frustrated Heisenberg antiferromagnet by differentiable symmetric tensor networks}

\author{Juraj Hasik}
\email{Juraj.hasik@irsamc.ups-tlse.fr}
\affiliation{
Laboratoire de Physique Th\'eorique UMR5152, C.N.R.S. and Universit\'e de Toulouse, \\
118 rte de Narbonne, 31062 Toulouse, FRANCE}

\author{Didier Poilblanc}
\affiliation{
Laboratoire de Physique Th\'eorique UMR5152, C.N.R.S. and Universit\'e de Toulouse, \\
118 rte de Narbonne, 31062 Toulouse, FRANCE}

\author{Federico Becca}
\affiliation{
University of Trieste, Department of Physics \\
Strada Costiera 11, 34151 Trieste, ITALY }

\date{\today}

\begin{abstract}
The recent progress in the optimization of two-dimensional tensor networks [H.-J. Liao, J.-G. Liu, L. Wang, and T. Xiang, Phys. Rev. X {\bf 9}, 
031041 (2019)] based on automatic differentiation opened the way towards precise and fast optimization of such states and, in particular, infinite 
projected entangled-pair states (iPEPS) that constitute a generic-purpose {\it Ansatz} for lattice problems governed by local Hamiltonians. In this 
work, we perform an extensive study of a paradigmatic model of frustrated magnetism, the $J_1-J_2$ Heisenberg antiferromagnet on the square lattice.
By using advances in both optimization and subsequent data analysis, through finite correlation-length scaling, we report accurate estimations of
the magnetization curve in the N\'eel phase for $J_2/J_1 \le 0.45$. The unrestricted iPEPS simulations reveal an $U(1)$ symmetric structure, which 
we identify and impose on tensors, resulting in a clean and consistent picture of antiferromagnetic order vanishing at the phase transition with 
a quantum paramagnet at $J_2/J_1 \approx 0.46(1)$. The present methodology can be extended beyond this model to study generic order-to-disorder 
transitions in magnetic systems.
\end{abstract}

\maketitle


\section{Introduction}\label{sec:level1}

The spin-$S$ antiferromagnet, with isotropic coupling $J_1$ between nearest-neighbor spins located on the sites of a square lattice, represents 
one of the most paradigmatic models of quantum magnetism. At zero temperature, the system develops long-range antiferromagnetic (N\'eel) order 
for any value of $S$: while for $S \ge 1$ there are analytical arguments~\cite{dyson1978,neves1986}, for the extreme quantum case with $S=1/2$, 
this has been numerically proven thanks to quantum Monte Carlo simulations on large systems~\cite{reger1988,sandvik1997,calandra1998}. Instead, 
any finite temperature will restore spin rotation symmetry, in agreement with the Mermin-Wagner theorem~\cite{mermin1966}. A magnetically
disordered ground state may be also achieved by including further super-exchange couplings, most notably a next-nearest-neighbor interaction
$J_2$, which destabilizes the N\'eel order driving towards a quantum phase transition. In this respect, much effort has been spent to understand 
the ground-state properties of the $J_1-J_2$ model defined by:
\begin{equation}
\label{eq:j1j2ham}
{\cal H} = J_1 \sum_{\langle i,j \rangle} \mathbf{S}_i \cdot \mathbf{S}_j + 
           J_2 \sum_{\langle\langle i,j \rangle\rangle} \mathbf{S}_i \cdot \mathbf{S}_j\, .
\end{equation}
Here, $\langle \dots \rangle$ and $\langle\langle \dots \rangle\rangle$ stand for nearest-neighbor and next-nearest-neighbor sites on the square 
lattice, respectively; $\mathbf{S}_i=(S^x_i,S^y_i,S^z_i)$ represents the spin-1/2 operator on the site $i$. Both the spin-spin interactions are
taken positive.

In the presence of finite $J_2$ a severe sign problem is present (especially in the local basis with $z$-component defined on each site), which 
prohibits quantum Monte Carlo algorithms from assessing large system sizes~\cite{sorella1998,choo2019,szabo2020}. Over the last three decades several alternative methods have been 
introduced and kept improving such as exact diagonalizations, density-matrix renormalization group (DMRG), functional-renormalization group (fRG), 
and variational Monte Carlo (VMC) approaches. The ground-state properties of the $J_1-J_2$ model have been intensively investigated, with 
contradicting results, supporting the existence of a valence-bond solid (with either columnar or plaquette order)~\cite{ed1,ed4,mambrini2006} 
or a spin liquid (either gapped or gapless)~\cite{jiang2012,mezzacapo2012,hu2013,hering2019}, or even 
both~\cite{gong2014,morita2015,wang2018,ferrari2020,nomura2020}. One important aspect emerging in the latest calculations is the existence of 
a {\it continuous} quantum phase transition between the antiferromagnetic and the paramagnetic phases for $J_2/J_1 \approx 0.5$, where the 
staggered magnetization (hereafter named simply ``magnetization'') goes to zero.

Recently, borrowing concepts from quantum information, tensor-network methods have been introduced~\cite{verstraete2004,verstraete2008,murg2009}.
In one dimension, the so-called matrix-product states (MPS) offer a convenient and elegant rephrasing of previous DMRG ideas. MPS evolved into 
the method of choice and provide very accurate approximations of the exact ground-state properties. Generalizations in two dimensions are more 
problematic. The prominent example, projected entangled-pair states (PEPS), provide the correct entanglement structure of most quantum ground 
states of local spin Hamiltonians~\cite{verstraete2006}, however, they suffer from a steep scaling of computational effort when enlarging the 
system size. For this reason, their application has been limited to ladder geometries with small number of legs~\cite{poilblanc2015,poilblanc2017a}
and finite 2D clusters with open boundary and up to $\approx 200-300$ sites~\cite{liu2018}. In order to overcome this computational barrier and 
avoid boundary effects, algorithms that work directly in the thermodynamic limit (dubbed iPEPS) have been introduced and 
developed~\cite{jordan2008,orus2009}: here, only a small number of tensors is explicitly considered and embedded into an environment that is 
self-consistently obtained (e.g., within the so-called corner-transfer matrix approaches~\cite{nishino1998} or channels~\cite{Vanderstraeten2016}).
The size of these tensors, and in turn the number of variational parameters of the wave function, is characterized by the so-called bond dimension 
$D$. The iPEPS are systematically improved by enlarging the bond dimension, accounting for increasingly entangled states. 

In recent years, iPEPS have been applied to assess the nature of the ground state of the $J_1-J_2$ model, mainly focusing on the highly-frustrated 
regime $J_2/J_1 \approx 0.5$~\cite{wang2013,poilblanc2017b,haghshenas2018}. However, these attempts were not completely satisfactory, since they 
either used a simplified tensor structure, limited to the description of paramagnets, or suffer from optimization problems, arising in methods 
that are not fully satisfactory and consistent (e.g., the so-called simple and full update~\cite{jiang2008,jordan2008}). In this respect, a 
breakthrough in the field has been achieved by performing the tensor optimization using the ideas of algorithmic differentiation, or better the 
{\it adjoint algorithmic differentiation} (AAD) technique, which allow a very efficient optimization even in the  presence of large number of  
parameters~\cite{liao2019}. Here, Liao and collaborators limited their application to the unfrustrated Heisenberg model (with $J_2=0$), showing 
that extremely accurate and completely stable results may be obtained for both the ground-state energy and magnetization.

Even though PEPS (and iPEPS) {\it Ans\"atze} are designed to describe both gapped and gapless states (following the entanglement entropy's area 
law, up to additive corrections), it remains an open question whether generic optimization can reliably reproduce highly-entangled ground states, 
as the ones that are possibly emerging in the frustrated regime $J_2/J_1 \approx 0.5$~\cite{hu2013,morita2015,ferrari2020,nomura2020}. Therefore, 
in this work, we do not directly address the question of the nature of the magnetically disordered phase; instead, we focus our attention to the 
magnetically ordered phase with $J_2/J_1 \le 0.45$ and perform an accurate determination of the magnetization curve as a function of the frustrating
ratio. In addition to its conceptual importance, the problem of the disappearance of antiferromagnetic order under increasing frustration offers a 
stringent test to most numerical methods, in general, and to tensor network methods, in particular. To this end, we apply the same ideas of AAD 
to optimize the iPEPS {\it Ansatz} for the $J_1-J_2$ model of Eq.~(\ref{eq:j1j2ham}). Importantly, unlike the previously proposed gradient-based 
optimizations~\cite{Vanderstraeten2016,corboz2016}, the AAD can be effortlessly extended beyond nearest-neighbour Hamiltonians. The energy and 
magnetization are obtained for different values of the bond dimension $D$, from $2$ up to $7$. Then, the estimates for $D \to \infty$ are obtained
for each frustration ratio $J_2/J_1$. Note however that, this is {\it not} realized by a crude extrapolation in $1/D$ (for which the results for 
different values of $D$ are considerably scattered) but, instead, by performing a correlation-length extrapolation, which is motivated by the 
finite-size scaling analysis that is well established in the N\'eel phase, as recently proposed in Refs.~\cite{rader2018,corboz2018}. Despite the 
fact that this mode of extrapolation requires the calculation of the correlation length $\xi$, which may not be as accurate as other thermodynamic 
quantities (e.g., energy and magnetization), it has been shown to give remarkably good results for the unfrustrated Heisenberg model. In fact, 
as we have mentioned earlier, even though iPEPS can describe certain gapless phases, their generic optimization instead leads to states with 
finite correlation lengths. e.g., in the N\'eel phase, and the bond dimension $D$ turn out not to be the correct object to quantify this aspect. 
As we will show, also in the presence of frustration, the analysis based on the correlation length gives reliable thermodynamic estimates, even 
though no exact results are available. Our calculations are compatible with a vanishing magnetization for $J_2/J_1 \approx 0.45$, which is in 
close agreement with recent calculations~\cite{hu2013,morita2015,wang2018,ferrari2020,nomura2020} and give a reference for future investigations.

The paper is organized as follows: in section~\ref{sec:method}, we will describe the iPEPS method; in section~\ref{sec:results}, we present the
results; in section~\ref{sec:concl}, we finally draw our conclusions and discuss the perspectives.

\section{iPEPS {\it Ansatz} and its optimization}\label{sec:method}

\subsection{General aspects}

We parametrize the state by a single real tensor 
\begin{equation}
    a^s_{uldr} =  \raisebox{-10pt}{\includegraphics[width=0.06\textwidth]{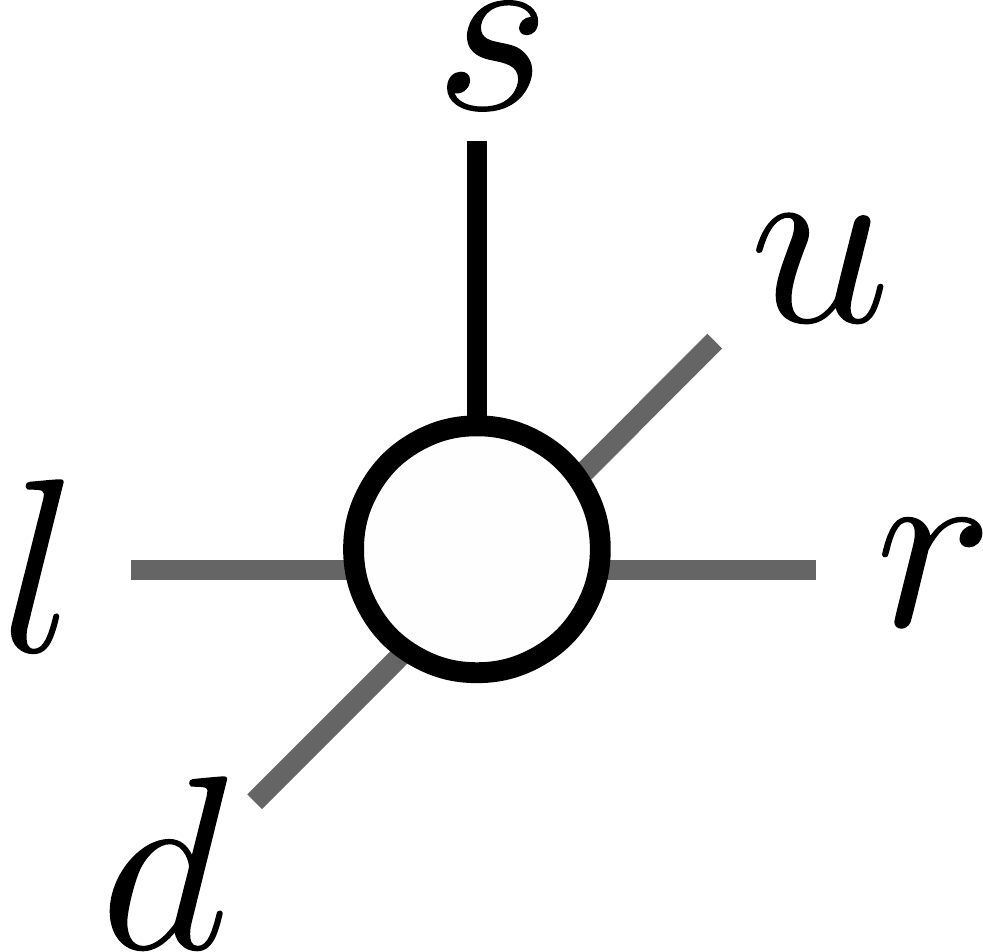}}
\end{equation}
with a physical index $s=\uparrow,\downarrow$ labeling the standard $S^z$ basis of the local physical Hilbert space and auxiliary (or virtual) 
indices $u,l,d,r$ of bond dimension $D$ (by convention running from $0$ to $D-1$ here). The physical wave function is then obtained by tiling 
the infinite square lattice with tensor $a$ and tracing over all auxiliary indices
\begin{gather}
\psi(a)= \sum_{\{s\}} c(a)_{\{s\}}|\{s\}\rangle \nonumber \\
c(a)_{\{s\}} := \text{Tr}_{aux}(a^{s_0}a^{s_1}a^{s_2}\ldots)
=
\raisebox{-18pt}{\includegraphics[width=0.15\textwidth]{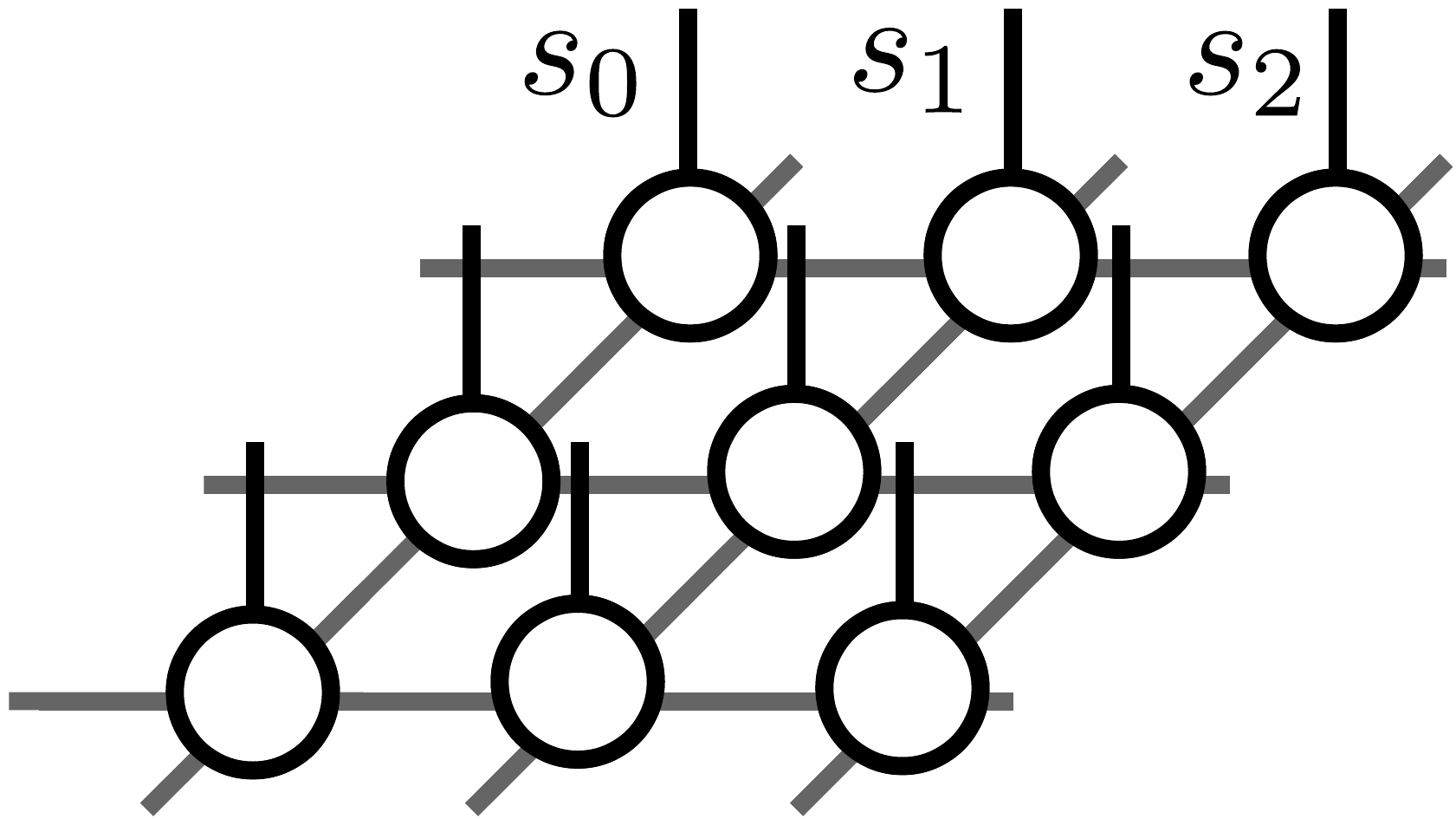}}
\end{gather}

The tensor $a$ is chosen (and constructed such as) to be invariant under a number of symmetries. First of all, it belongs to the $A_1$ irreducible 
representation of the $C_{4v}$ point group, thus enforcing all the spatial symmetries of the square lattice on the iPEPS. The antiferromagnetic 
correlations are incorporated in the ansatz by unitaries $-i\sigma^y$, which rotate the physical $S^z$ basis at every site of one sublattice. 
We absorb these unitaries into observables leaving the definition of the wave function untouched [see Eq.~(\ref{eq:epersite})]. 

Secondly, the tensor $a$ also possesses a further structure by requiring certain transformation properties under the action of $U(1)$ group (see 
below). Such choice is motivated by the remaining $U(1)$ symmetry in the ordered phase, which manifests itself as equivalence between different 
magnetizations connected by transverse (Goldstone) modes. As defined below, $U(1)$ tensor classes are defined by assigning specific ``charges'' 
to the virtual and physical degrees of freedom.

When considering $A_1$- and $U(1)$-symmetric states, the tensor $a=a(\vec{\lambda})$ is taken to be a linear combination of (fixed) elementary 
tensors $\{t_0, t_1, \ldots \}$ (named a tensor ``class'') such that 
\begin{equation}
a(\vec{\lambda}) = \sum_i \lambda_i t_i  \, ,
\end{equation}
with coefficients $\vec{\lambda}$ being the variational parameters. The elementary tensors $\{t_0, t_1, \ldots \}$ are different representatives 
of the $A_1$ irreducible representation for some choice of the $U(1)$ charges.

Given an iPEPS defined by tensor $a$, the evaluation of any observable $\mathcal{O}$ amounts to a contraction of infinite double-layer network 
composed of tensors $a$ together with the tensor representation of $\mathcal{O}$. Such tensor network is the diagrammatic equivalent of usual 
expression $\langle \mathcal{O} \rangle = \langle\psi(a)| \mathcal{O} |\psi(a)\rangle$. A central aspect of iPEPS method is an approximate 
contraction of such networks. In this work, we realize them by finding the so-called environment tensors $C$ and $T$ of dimension $\chi$, dubbed 
environment dimension, by the means of corner-transfer matrix (CTM) procedure~\cite{nishino1998}. These tensors compress the parts of the original 
infinite network in approximate but finite-dimensional objects. Afterwards, the desired reduced density matrices can be constructed from $C$ and
$T$, together with the on-site tensor $a$. Ultimately, the exact value of any observable is recovered taking $\chi \to \infty$, which 
we extrapolate from the data for increasingly large $\chi$.

The optimization of tensor $a$ (or equivalently in the $U(1)$-symmetric approach, $\vec{\lambda}$) is carried out using standard gradient-based 
method L-BFGS supplemented with backtracking linesearch. The gradients are evaluated by AAD, which back-propagates the gradient through the 
whole process of energy evaluation for fixed $\chi$~\cite{liao2019}: Starting with a given CTM, followed by assembling the reduced-density 
matrices from converged ${C,T}$ tensors and finally evaluating the spin-spin interaction between nearest and next-nearest neighbors. 

\begin{figure}
\includegraphics[width=\columnwidth]{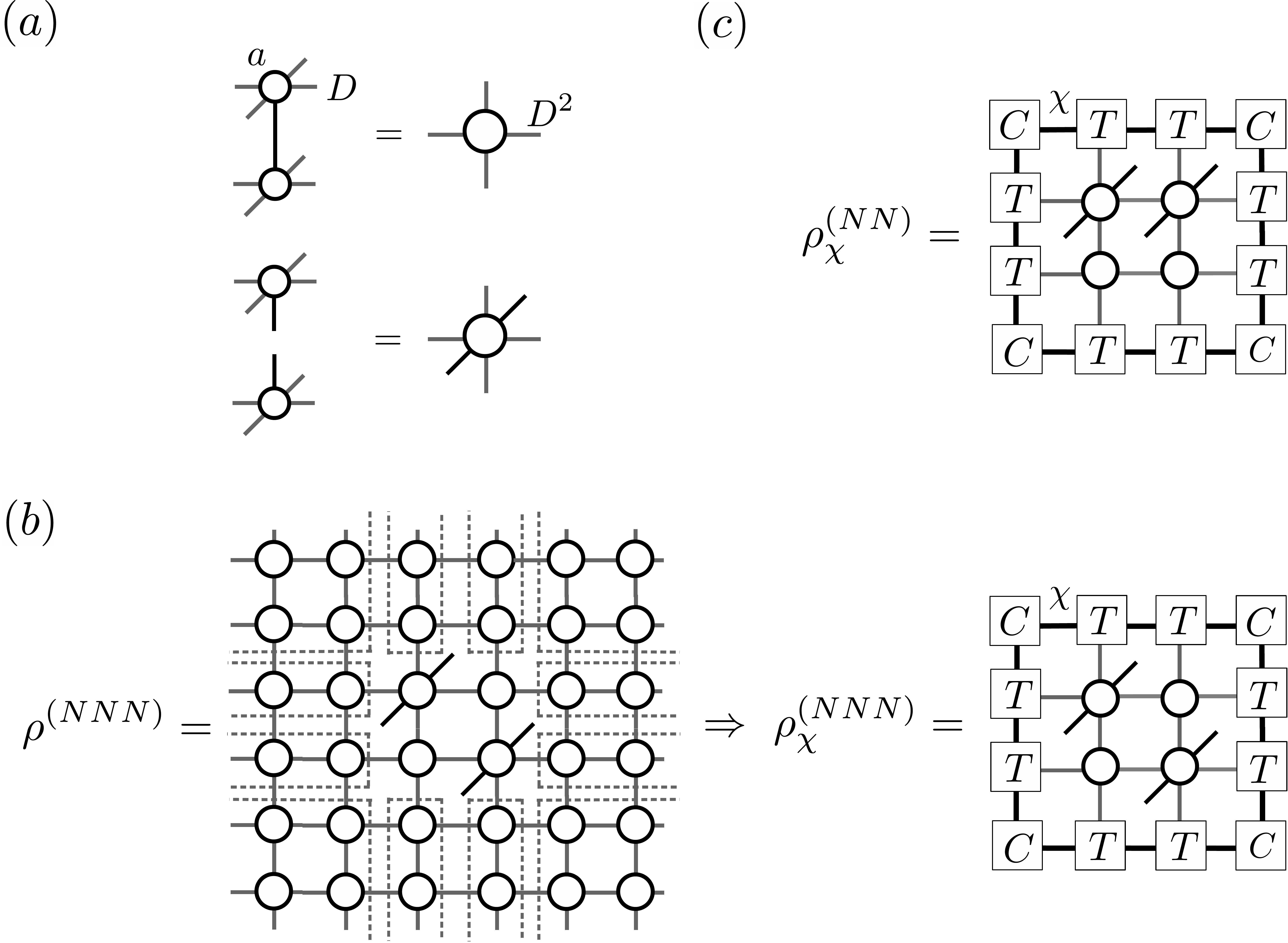}
\caption{\label{fig:rdms}
Definition of reduced-density matrices necessary for evaluating the energy per site of the $J_1-J_2$ model over single-site iPEPS with 
$C_{4v}$ symmetry. (a) Double-layer tensors with contracted and uncontracted physical indices. (b) Infinite tensor network corresponding to the
next-nearest-neighbour $\rho^{(NNN)}$ as approximated by $\rho^{(NNN)}_\chi$ in the finite network with $C$ and $T$ tensors resulting from CTM. 
(c) Finite-network approximation of nearest-neighbour $\rho^{(NN)}_\chi$ within the same $2\times2$ cluster.}
\end{figure}

\begin{figure}
\includegraphics[width=\columnwidth]{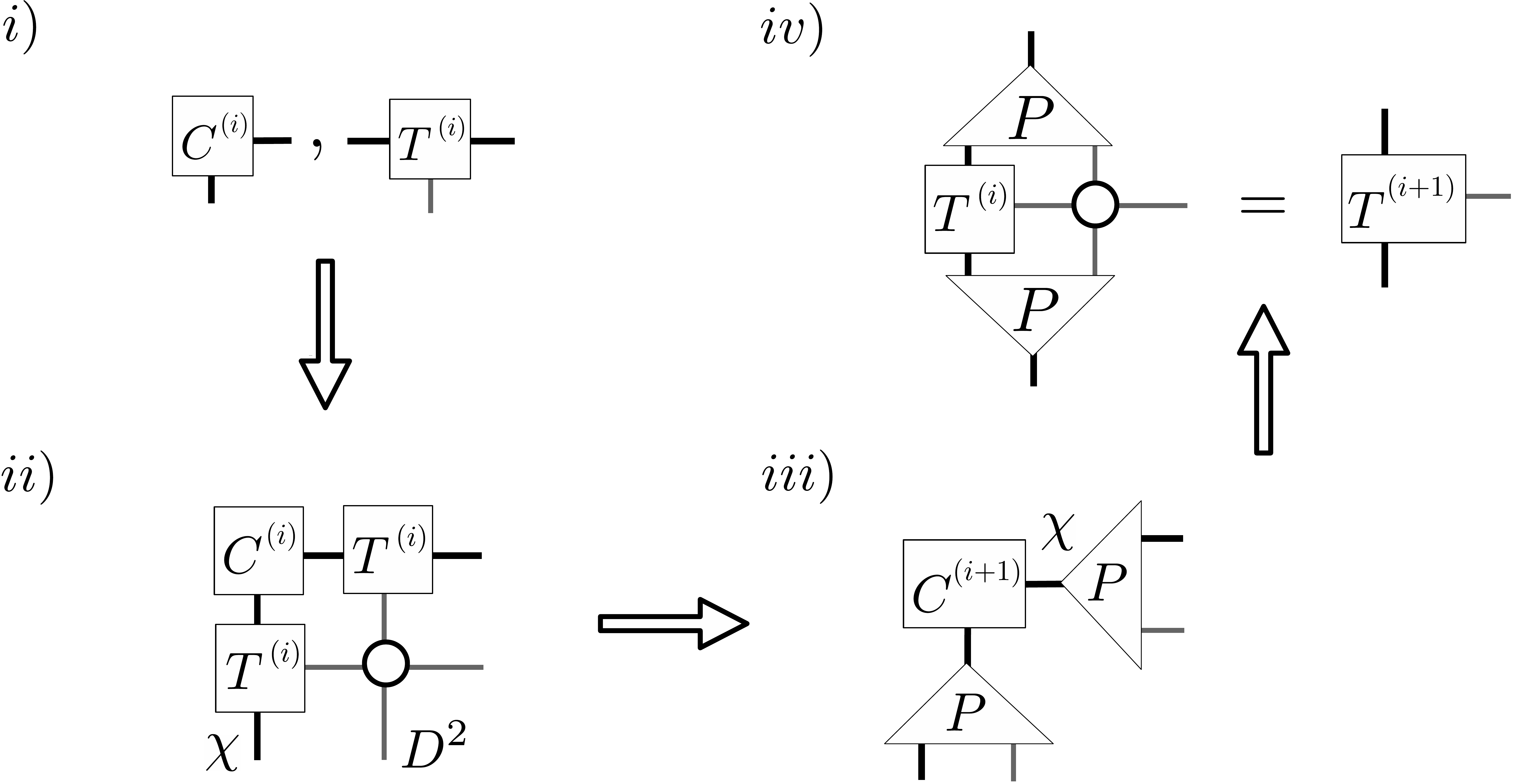}
\caption{\label{fig:ctm}
Key steps of the CTM algorithm for single-site iPEPS with $C_{4v}$ symmetry. (i) Initial tensors at iteration $i$: $\{ C^{(i)}, T^{(i)} \}$. 
(ii) Construction of enlarged corner and its reshaping into matrix of dimensions $D^2\chi \times D^2\chi$. (iii) Symmetric eigenvalue decomposition 
of enlarged corner and truncation down to leading $\chi$ eigenpairs by magnitude of the eigenvalues. Truncation is always done at the boundaries 
between degenerate eigenvalues (see text). (iv) Absorption and truncation with isometry $P$ from step (ii) for half-row/-column tensor $T$.}
\end{figure}

\subsection{Extracting the relevant $U(1)$ charges}\label{subsec:smith}

For small enough frustration, in the N\'eel phase, the unconstrained optimization of tensor $a$ leading to correct $U(1)$-symmetric iPEPS would be 
a desirable outcome. Under circumstances, AAD optimization can arrive at an almost $U(1)$-symmetric tensor $\tilde{a}$. In such case, a direct and 
robust evidence can be seen in the nearly degenerate pairs of leading eigenvalues of the transfer matrix. Importantly, such iPEPS states provide 
an unbiased information about the energetically favourable $U(1)$-charge structure of tensor $a$. We are concerned with inferring these charges 
from the elements of tensor $\tilde{a}$. Obtaining the correct charge assignment of the smallest $D$ tensors allows (i) to perform an efficient 
variational optimization over a greatly reduced number of parameters $\vec{\lambda}$, (ii) to obtain truly $U(1)$-symmetric environments via CTM 
and, finally, (iii) to predict the correct charge content of higher-$D$ $a$ tensors and, hence, enable to perform (i) and (ii) for larger $D$. 

Before describing how to achieve the goal of obtaining the charges from the almost symmetric $\tilde{a}$ tensor, let us first briefly review the 
expected properties of the resulting $U(1)$-symmetric $a$ tensor. In practice, one has to assign $U(1)$ charges $\vec{u}=(u^\uparrow, u^\downarrow)$
and $\vec{v}=(v_0,\ldots,v_{D-1})$ to the two physical spin-1/2 components and the $D$ virtual degrees of freedom on each of the four auxiliary indices. Without loss of generality we take them to be integers. In this language, $U(1)$ invariance is realized by simply enforcing a selection rule for the non-zero tensor
elements $a^{s}_{uldr}$ which should exhibit a local charge conservation
\begin{equation}
\label{eq:u1symcond}
u^{s} + v_{u} + v_{l} + v_{d} + v_{r} = N\, ,
\end{equation}
where $N$ is some fixed integer. Notice that in order to preserve  $C_{4v}$ symmetry the same $\vec{v}$, associated to all the virtual indices, 
is taken on the four legs of the tensor. Note also that there is some freedom in the definition of the charges since shifts like 
$u^s \to u^s+\alpha$, $v_\sigma \to v_\sigma+\beta$, and $N \to N+\alpha+4\beta$, with $\alpha$ and $\beta \in \mathds{Z}$, leaves 
Eq.~(\ref{eq:u1symcond}) invariant. It is easy to connect this charge conservation to the $U(1)$ invariance of the $a$ tensor. Indeed, the 
action of any element $g \in U(1)$ on $a$ is given by:
\begin{equation}
a^s_{uldr} \to (ga)^s_{uldr} = a^{s'}_{u'l'd'r'} U^{ss'} V_{uu'} V_{ll'} V_{dd'} V_{rr'},     
\end{equation}
where $U$ and $V$ are diagonal matrices depending on $g$, and all auxiliary indices are transformed by the same $V$:
\begin{eqnarray}
U^{ss'} &=& e^{i\theta_g u^s}\delta^{ss'}, \\
V_{\gamma \gamma'} &=& e^{i\theta_g v_{\gamma}}\delta_{\gamma \gamma'},
\end{eqnarray} 
with the phase $\theta_g \in \mathbb{R}$ and $\gamma=0,\ldots,D-1$. Therefore, the non-zero elements of the tensor $a$ transform according to 
\begin{equation}\label{eq:u1trans}
(ga)^s_{uldr} = a^{s}_{uldr} e^{i\theta_g(u^{s} + v_{u} + v_{l} + v_{d} + v_{r})} 
\end{equation}
Hence, Eq.~(\ref{eq:u1symcond}) implies that $a$ is indeed invariant up to global phase under the action of $U(1)$. Once the relevant $U(1)$ charges $\vec{u}$ and 
$\vec{v}$ are known (see below), practically, Eq.~(\ref{eq:u1symcond}) is used in the construction of the elementary tensors $\{t_0, t_1, \dots \}$ 
by filtering out their non-zero elements.

\begin{table}[b]
\caption{\label{tab:table1}
$U(1)$ charges as inferred from unrestricted simulations with bond dimensions $D=2,\dots,7$. Predictions of the charges for $D=8$ and $9$ are 
also shown. Note that the ordering of the $v_{\alpha}$ charges is arbitrary and the gauge freedom has been fixed by taking $N=1$. The last column 
shows the number of elementary tensors $t_i$. }
\begin{ruledtabular}
\begin{tabular}{lll}
$D$ & $[ u_\uparrow, u_\downarrow, v_0,v_1,\cdots, v_{D-1}] $  & number of tensors\\
\colrule
2 & $[1,-1,0,2]$ & 2 \\
3 & $[1,-1,0,2,0]$& 12 \\
4 & $[1,-1,0,2,-2,0]$& 25 \\
5 & $[1,-1,0,2,-2,0,2]$& 52 \\
6 & $[1,-1,0,2,-2,0,2,-2]$&  93 \\
7 & $[1,-1,0,2,-2,0,2,-2,2]$&  165 \\
8 & $[1,-1,0,2,-2,0,2,-2,0,2]$& 294 \\
9 & $[1,-1,0,2,-2,0,2,-2,0,2,-2]$& 426 \\
\end{tabular}
\end{ruledtabular}
\end{table}

Let us now describe how to infer the charges from an unrestricted tensor optimization that has produced an almost symmetric on-site tensor 
$\tilde{a}$, with bond dimension $D$. To identify the dominant (at least for small $D$) $U(1)$-symmetric component of $\tilde{a}$, and then 
ultimately derive the hidden $U(1)$ charges, we have to first perform a higher-order singular value decomposition of $\tilde{a}$:
\begin{equation}
{\tilde a}^s_{uldr} = Z^{ss'}Y_{uu'}Y_{ll'}Y_{dd'}Y_{rr'} c^{s'}_{u'l'd'r'},
\end{equation}
with unitary matrices $Z$, $Y$, and the so-called core tensor $c$. The same unitary $Y$ is associated to different auxiliary legs due to the 
enforced $C_{4v}$ symmetry. The core tensor $c$ plays an analogous role to singular values in standard singular value decomposition of a matrix. 
The untruncated core tensor $c$ by itself defines a physically equivalent iPEPS to the one given by $\tilde{a}$. A good lower-rank approximation 
of $\tilde{a}$ can be obtained by truncation of the smallest elements of the core tensor $c$.  The basic premise, supported by nearly degenerate 
transfer matrix spectrum for small $D$, is that the relative magnitude of symmetry breaking elements of tensor $c$ is small. Therefore, we assume 
that the largest elements of tensor $c$ respect the $U(1)$-symmetry constrain associated to an unknown set of charges $\vec{u}$ and $\vec{v}$. 

For the last step in identifying the charges, we re-formulate the problem in terms of linear algebra. First, taking a set of $n$ largest tensor 
elements (modulo $C_{4v}$ symmetry), and writing down Eq.~(\ref{eq:u1symcond}) for each of them will result in a set of $n$ coupled linear 
equations (with integer coefficients) of the $D+2$ unknown charges. Whenever $n>D+2$, the linear system becomes over-complete and, increasing $n$ 
still allows the same solution for the charges, unless $n$ is taken too large so that (small) non-zero tensor elements breaking $U(1)$-symmetry 
are included.  To solve this linear problem it is convenient to recast the constraints into a $n \times (D+2)$ matrix. The matrix, containing 
integer matrix elements, is obtained by simply counting the total number of charges of each type $\gamma$ and $s$ on the virtual and physical 
legs, respectively. More precisely, we define vectors $\vec{n}(c^s_{uldr})$ of integer coordinates that count the number of times specific 
{\it index value} appears among the indices of a given tensor element. Expressing each individual element constraint~(\ref{eq:u1symcond}) as 
$\vec{n}(c^s_{uldr})\cdot(\vec{u},\vec{v}) = N$ and recasting them into matrix form, the linear system can be written in a compact fashion as 
$M\cdot(\vec{u},\vec{v}) = \vec{N}$. 

To be explicit, let us consider the case $D=3$ for which all charges can be obtained using only the $n=D+2=5$ largest tensor elements of tensor $c$:
\begin{equation}
\begin{matrix}
\vec{n}(c^\uparrow_{0000}) & \to \\
\vec{n}(c^\downarrow_{0001}) & \to \\
\vec{n}(c^\uparrow_{0002}) & \to \\
\vec{n}(c^\uparrow_{2222}) & \to \\
\vec{n}(c^\uparrow_{0222}) & \to
\end{matrix}
\begin{bmatrix}
1 & 0 & 4 & 0 & 0 \\
0 & 1 & 3 & 1 & 0 \\
1 & 0 & 3 & 0 & 1 \\
1 & 0 & 0 & 0 & 4 \\
1 & 0 & 1 & 0 & 3 
\end{bmatrix}
\cdot
\begin{bmatrix}
u^\uparrow \\
u^\downarrow \\
v_0 \\
v_1 \\
v_2 
\end{bmatrix}
=
\begin{bmatrix}
N \\
N \\
N \\
N \\
N
\end{bmatrix}.
\label{eq:matrixM}
\end{equation}
If the tensor $c$ possesses an (approximate) $U(1)$ symmetry structure (as in the example above), then the linear system has a non-trivial solution 
in terms of charges $\vec{u}$ and $\vec{v}$. To solve it, it is known that one needs to bring the matrix $M$ into its Smith normal form (see 
Appendix~\ref{appA}). Note here that the integer $N$ can, in fact, be changed arbitrarily. Although, the explicit values of the charges will depend 
on $N$, the $U(1)$ class of $a$ tensors will not. In other word, there is some ``gauge'' freedom to determine each $U(1)$ class. For the example
with $D=3$ considered here, we get integer charges, $u^\uparrow=+1$, $u^\downarrow=-1$, $v_0=0$, $v_1=2$ and $v_2=0$, as can be checked by direct 
substitution in Eq.~(\ref{eq:matrixM}) choosing $N=1$. A complete list of the relevant charges are shown in Table~\ref{tab:table1} for bond 
dimension up to $D=9$. 

\subsection{Reduced density matrices, CTM algorithm, and implementation details}

The evaluation of energy is realized through two distinct reduced-density matrices (RDM), $\rho^{(NN)}$ and $\rho^{(NNN)}$, for nearest and 
next-nearest neighbour sites respectively. Their diagrammatic definition is shown in Fig.~\ref{fig:rdms}. The energy per site is then given by:
\begin{equation}
\label{eq:epersite}
e = 2 J_1 \text{Tr} \left [ \rho^{(NN)} \mathbf{S} \cdot \mathbf{\tilde{S}} \right ] + 
    2 J_2 \text{Tr} \left [ \rho^{(NNN)} \mathbf{S} \cdot\mathbf{S} \right ],  
\end{equation}
with $\tilde{S}^\alpha = -\sigma^y S^\alpha (\sigma^y)^T$, as these are the only non-equivalent terms of Hamiltonian~(\ref{eq:j1j2ham}) acting 
on the single-site iPEPS with $C_{4v}$ symmetry.

The two RDMs are obtained by substituting the environment of a $2\times2$ cluster within the infinite tensor network with the CTM approximation 
and tracing out all but two (nearest-neighbor or next-nearest-neighbor) sites. The leading computational cost in contraction of these networks is 
$O[(\chi D^2)^3 p^2]$ with $p=2$ being the dimension of the physical index $s$. A more complete alternative is to consider a RDM of {\it all} four 
spins contained within the cluster. However, contracting such network with eight open physical indices is more expensive in terms of computational 
complexity and memory requirements, as both are amplified by a factor of $p^2$.

The most demanding part of the calculations is the CTM algorithm. Given the highly constrained nature of our iPEPS, in particular the $C_{4v}$ 
symmetry imposed on tensor $a$, we can utilize the efficient formulation of the algorithm of Ref.~\cite{nishino1998}. The $C_{4v}$ symmetry 
of the on-site tensor $a$ is reflected in the corner matrix $C$ which is taken to be diagonal and half-row/-column tensor $T$ which is symmetric 
with respect to the permutation of its environment indices. We show the diagrammatic description of the main steps within single CTM iteration 
in the Fig.~\ref{fig:ctm}. 

There are few more remarks to be made regarding the implementation of the CTM algorithm. The initial $C$ and $T$ tensors are given by partially 
contracted double-layer tensor, e.g. $C_{(dd')(rr')}=\sum_{sul} a^s_{uldr}a^s_{uld'r'}$. In addition, after each step of the CTM the tensors $C$
and $T$ are symmetrized accordingly and normalized by their largest element. To establish the convergence of the CTM, we use the RDM of nearest 
neighbors $\rho^{(NN)}_{2\times1}$ computed just from the $2\times1$ cluster at each CTM step. Once the difference (in sense of Frobenius norm) 
between $\rho^{(NN)}_{2\times1}$ from two consecutive iterations becomes smaller than $\epsilon_{CTM}$, we consider the CTM converged. During 
optimization we set $\epsilon_{CTM} = 10^{-8}$, which typically requires at most $O(70)$ iterations to converge for largest $(D,\chi_{opt})=(7,147)$
simulations considered. For scaling of observables of optimized states we instead iterate CTM until $\epsilon_{CTM} = 10^{-12}$. Remarkably, the 
$U(1)$ symmetry is preserved along the CTM procedure, whenever we adjust the truncation as to never break the multiplet structure of the enlarged 
corner.

Finally, a peculiar complication is present in the process of computing gradients by AAD, with two distinct aspects. First, the standard definition 
of adjoint function of eigenvalue (or singular value) decomposition relies on computing the full decomposition~\cite{giles2008}. Hence, in this 
context one cannot resort to significantly faster partial decompositions such as Lanczos (at least during gradient computation). This sets the 
leading complexity of CTM iteration to $O[(\chi D^2)^3]$. Recently, a developed differentiable dominant eigensolver tries to address this 
shortcoming by alternative adjoint formula~\cite{Xie2020}. The second, more fundamental aspect is the ill-defined adjoint in the case of degenerate 
eigenvalues stemming from the terms proportional to the inverse of spectral gaps. We use a smooth cutoff function~\cite{liao2019} to tame this 
problematic terms. Although doing so, the accidental crossings of eigenvalues in course of CTM sometimes result in erroneous gradients. In general, 
we found this occurrence, manifested by the failure of linesearch, to be rare. The formulation of AAD applied to gauge-invariant scalars (such as 
energy), whose computation however involves eigendecomposition with degenerate spectrum, still remains an open problem.

The complete algorithm is available as a part of the open-source library {\it peps-torch}~\cite{pepstorch} focused on AAD optimization of iPEPS.

\begin{figure}
\includegraphics[width=\columnwidth]{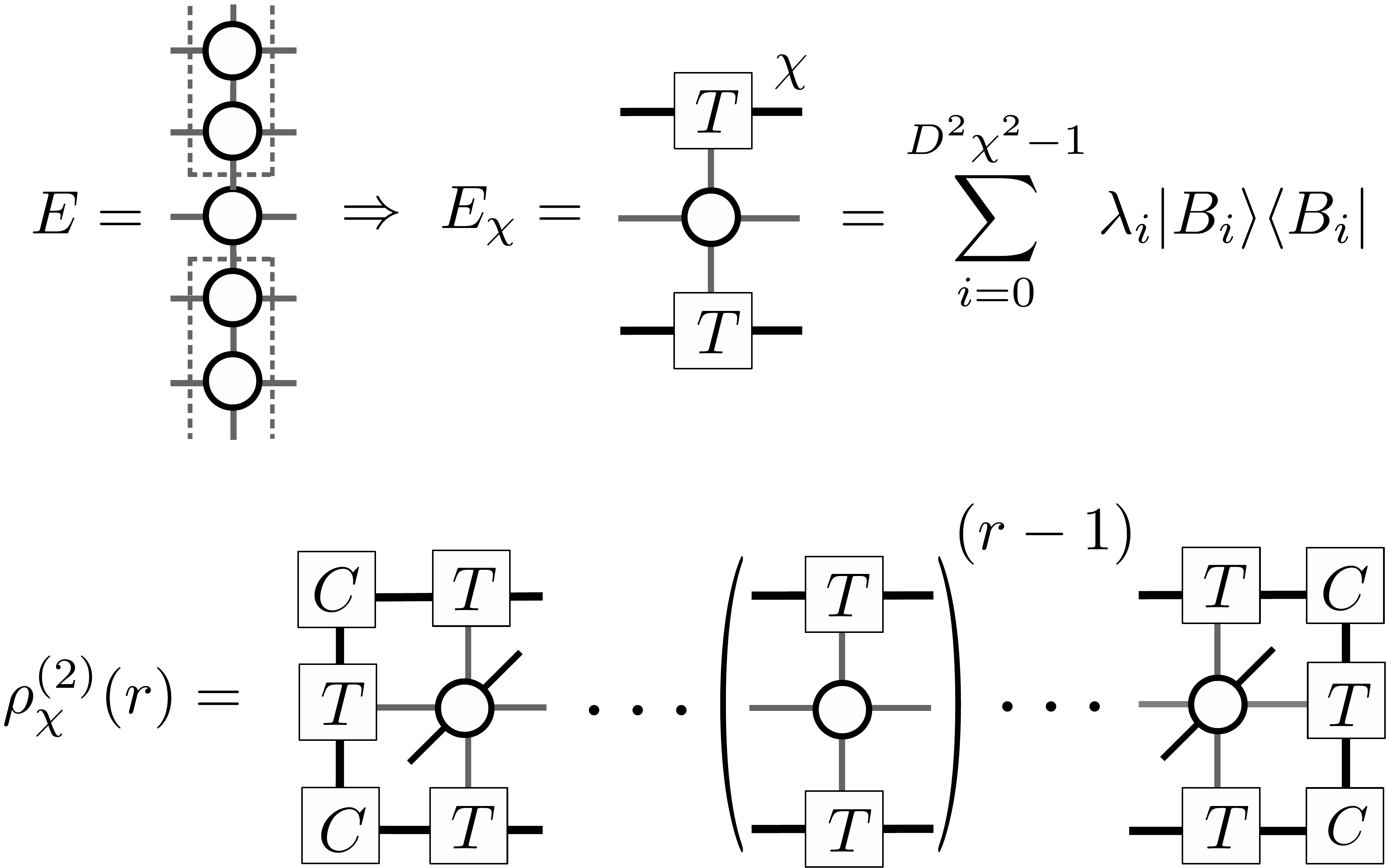}
\caption{\label{fig:TM} 
Top: Definition of the transfer matrix $E$ and its finite-$\chi$ approximation $E_\chi$ given by the converged $T$ tensor. Due to $C_{4v}$ symmetry 
imposed on the ansatz, the transfer matrix is symmetric and can be diagonalized. Eigenvalues are ordered with descending magnitude with the leading
eigenvalue $\lambda_0$ normalized to unity. Bottom: RDM for two-point correlation functions, defined for $r \geq 1,$ and its connection to transfer 
matrix $E$.}
\end{figure}

\begin{figure*}
\includegraphics[width=\textwidth]{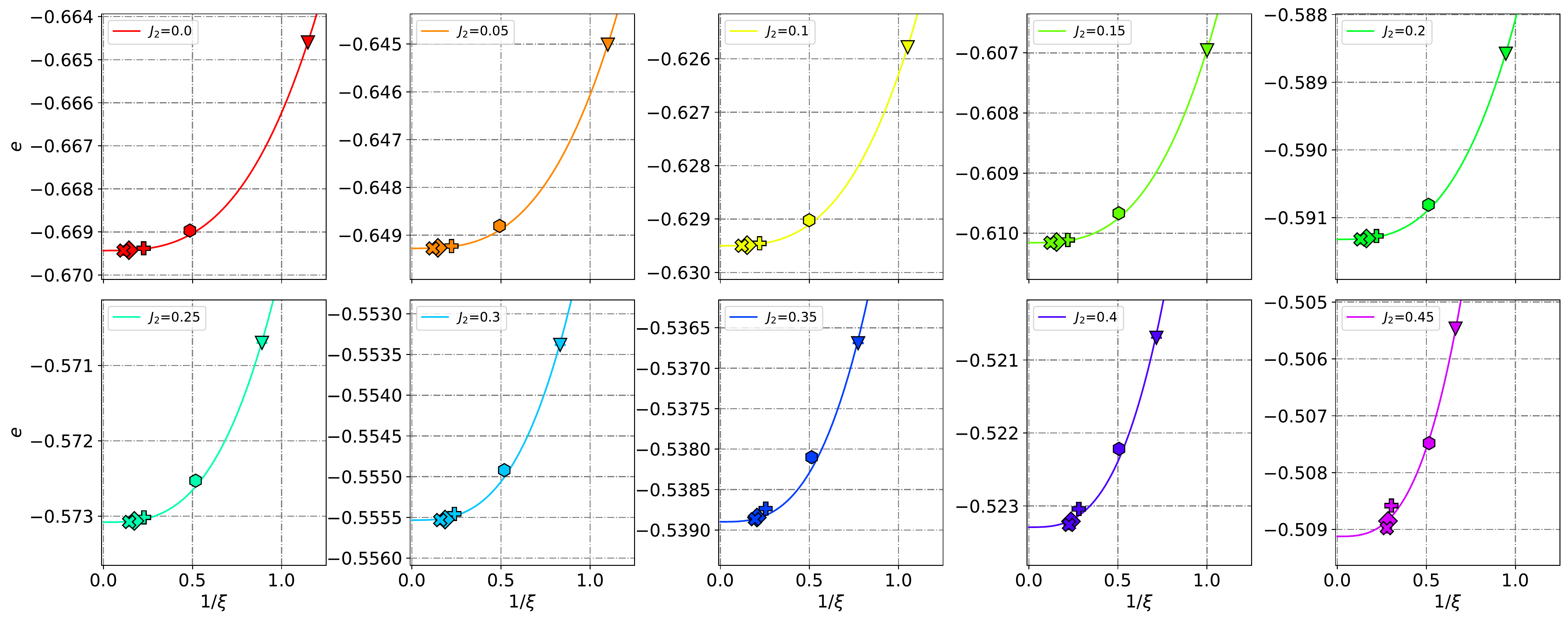}
\caption{\label{fig:energy} 
Finite correlation-length scaling of the energy per site for the $C_{4v}$-symmetric $U(1)$ iPEPS {\it Ansatz} with bond dimensions $D=3,\dots,7$ 
(denoted by triangles, hexagons, pluses, diamonds, and crosses in the same order). 
Continuous lines are linear fits in $1/\xi^3$ which is the expected scaling in the magnetically ordered phase~\cite{rader2018}.} 
\end{figure*}

\begin{figure*}
\includegraphics[width=\textwidth]{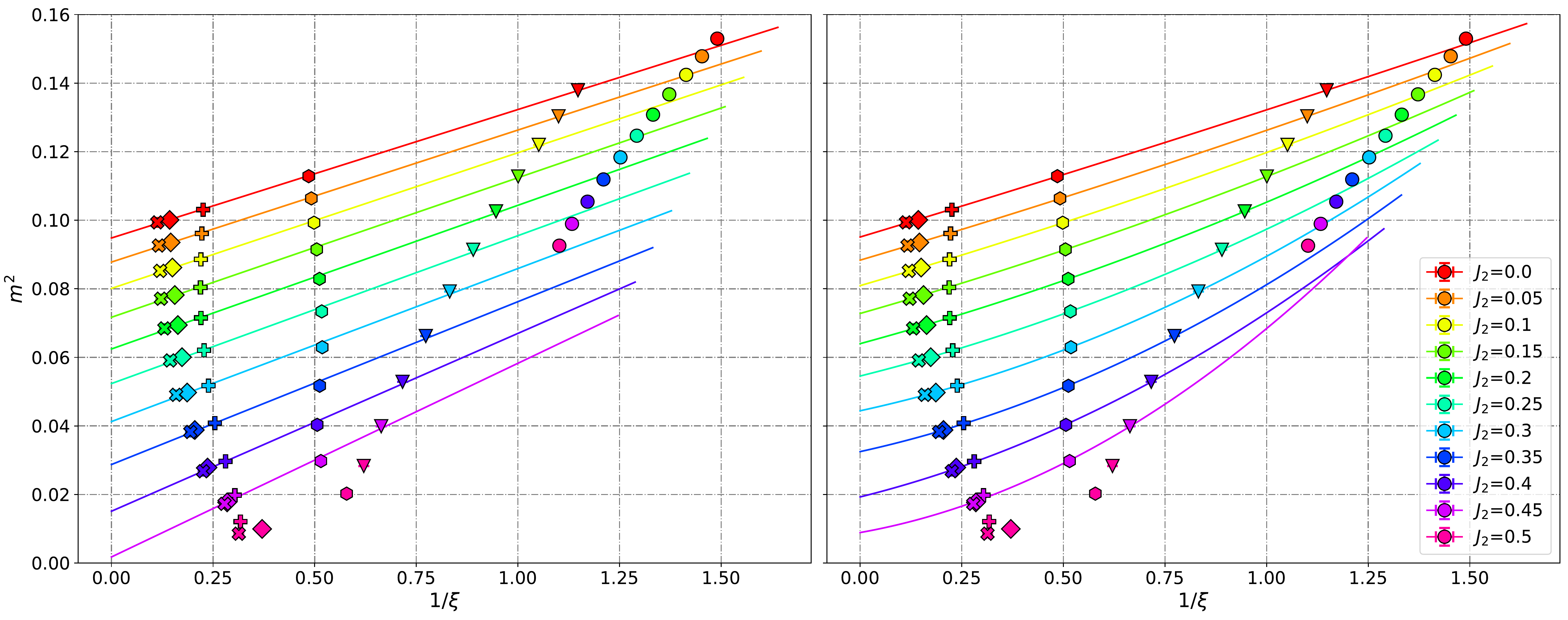}
\caption{\label{fig:magnet}
Finite correlation-length scaling of the magnetization for the $C_{4v}$-symmetric $U(1)$ iPEPS {\it Ansatz} with bond dimensions $D=2,\dots,7$
(denoted by circles, triangles, hexagons, pluses, diamonds, and crosses in the same order). The magnetization is plotted as a function of $1/\xi$, 
expected in the magnetically ordered phase~\cite{rader2018}. Linear (quadratic) extrapolations of magnetization, excluding $D=2$ data, are reported 
in the left (right) panel, except for $J_2/J_1=0.5$.}
\end{figure*}

\section{Results}\label{sec:results}

Our analysis is based upon an extensive set of calculations for various bond dimensions, ranging from $D=2$ to $7$, and different values of the
frustrating ratio $J_2/J_1$ up to $0.5$. For the large bond dimensions considered, the optimizations have been performed with environment dimensions 
up to $\chi_{opt}=4D^2$ in the case of $D=5,6$ and up to $\chi_{opt}=3D^2$ for $D=7$. Here, we want to highlight a few important aspects of iPEPS 
that are crucial for the investigation of the magnetically ordered phase. First of all, within optimizations with no imposed symmetries, there is 
a generic tendency to break the physical $U(1)$ symmetry of the N\'eel state (corresponding to global rotations around the axis of the spontaneous 
magnetization), leading to a slight (spin) nematic order, e.g., different values of the nearest-neighbour $S^xS^x$ and $S^yS^y$ correlations. This 
effect becomes more severe with increased frustration. For example, for most of the states with $D>3$ and $J_2 \gtrsim 0.3$, there is a sensible 
(e.g., $5-10\%$ and even larger) difference in the correlation lengths corresponding to the transverse directions. Connected to this issue, we 
observe that it is possible to stabilize distinct ``families'' of local minima for various bond dimensions $D$, in particular $D=3$ and $4$, with 
substantial differences in their magnetization, correlation length, and the degree of nematic order. Every family corresponds to a specific way 
the quantum fluctuations are built on top of the classical N\'eel state, e.g., by converging towards one of the possible choices of $U(1)$ charges 
or breaking the symmetry completely. Given the limited number of bond dimensions that are available within our AAD optimization, it is then of 
utmost importance to identify the family of minima that are connected and lead to a smooth and physically sound extrapolation in the $D \to \infty$ 
limit. Therefore, using the scheme introduced in Sec.~\ref{subsec:smith}, we take the optimized and almost $U(1)$-symmetric states from unrestricted 
simulations (typically for $J_2 \approx 0$) and infer their charge structure. The charges revealed by this analysis are listed in 
Table~\ref{tab:table1} and define the correct classes of $C_{4v}$-symmetric $U(1)$ iPEPS for $D$ ranging from $2$ to $7$, which best describe 
the N\'eel phase.

In order to obtain the thermodynamic estimates of the ground-state energy and magnetization (within the magnetically ordered phase), we compute 
these quantities for increasing values of the bond dimension $D$. A brute-force extrapolation in $1/D$ provides poor estimates, given the fact 
that the data are usually scattered, see for example the case of the magnetization reported in Appendix~\ref{appB}. Instead, we follow the recent 
proposal that has been put forward in Refs.~\cite{rader2018,corboz2018}. In this respect, for every value of $D$ used, we compute the dominant 
correlation length $\xi$ which is defined by the so-called transfer matrix $E$ of iPEPS, see Fig.~\ref{fig:TM}:
\begin{equation}
\xi = - \frac{1}{\log|\lambda_1|},
\end{equation}
where $\lambda_1$ is the second largest eigenvalue of the transfer matrix (without the loss of generality we assume that the largest one is 
normalized to 1). We remark that the value of $\xi$ obtained in this way coincides with the correlation length of the usual spin-spin correlation 
function (or, more precisely, the transverse correlations):
\begin{equation}\label{eq:spinspin}
\langle\mathbf{S}_0\cdot\mathbf{S}_r\rangle = 
\begin{cases}
\text{Tr}[\rho^{(2)}(r)\mathbf{S}\cdot\mathbf{S}] & r \in {\rm even} \\
\text{Tr}[\rho^{(2)}(r)\mathbf{S}\cdot\mathbf{\tilde{S}}] & r \in {\rm odd}
\end{cases},
\end{equation}
where $\rho^{(2)}(r)$, defined in Fig.~\ref{fig:TM}, is the two-point RDM. To obtain the $\chi \to \infty$ limit of the correlation length, we 
use the scaling formula~\cite{rams2018,rader2018}:
\begin{equation}
\frac{1}{\xi(\chi)} = \frac{1}{\xi(\infty)} + \alpha \left(\log \left| \frac{\lambda_3(\chi)}{\lambda_1(\chi)} \right|\right)^\beta,
\end{equation}
which allows for more precise extrapolation of $\xi$ than the usual $1/\chi$ scaling across all ratios of $J_2/J_1$~\footnote{In general one uses 
ratio of the second and third largest eigenvalues, $\lambda_1$ and $\lambda_2$; however, due to $U(1)$ symmetry, they are always degenerate,
forcing us to consider the next largest eigenvalue $\lambda_3$.}.

Finally, the thermodynamic estimates of the energy and magnetization (squared) are obtained by a suitable fit in powers of $1/\xi$:
\begin{eqnarray}
e(\xi) &=& e(\infty) + \frac{A}{\xi^3} + O \left(\frac{1}{\xi^4}\right), \\
m^2(\xi) &=& m^2(\infty) + \frac{B}{\xi} + O \left(\frac{1}{\xi^2}\right)\, ,
\end{eqnarray}
where $m=|\text{Tr}[\rho^{(1)}\mathbf{S}]|$ and $\rho^{(1)}$ is the single site RDM.
\begin{table}
\caption{\label{tab:enemag}
Ground-state energies (in units of $J_1$) $e(D,\chi)$ and magnetization square $m^2(D,\chi)$ for $D=7$, which can be considered as upper bounds 
of the exact $D \to \infty$ values. The tensor was optimized up to an environment dimension $\chi_{opt}=3D^2=147$. The $\chi \to \infty$ 
extrapolations are done from  environment bond dimensions $\chi \in [D^2,13D^2]$.}
\begin{ruledtabular}
\begin{tabular}{lcccc}
$J_2/J_1$ & $e(7,147)$ & $e(7,\chi \to \infty )$ & $m^2(7,147)$ & $m^2(7,\chi \to \infty )$ \\
\hline
0.0 & -0.669428 & -0.669432 &  0.0994 & 0.0994 \\
0.05 & -0.649273 & -0.649277 &  0.0926 & 0.0926 \\
0.1 & -0.629497 & -0.629501 &  0.0852 & 0.0852 \\
0.15 & -0.610154 & -0.610159 &  0.0771 & 0.0771 \\
0.2 & -0.591314 & -0.591320 &  0.0685 & 0.0685 \\
0.25 & -0.573067 & -0.573076 &  0.0591 & 0.0591 \\
0.3 & -0.555520 & -0.555533 &  0.0491 & 0.0491 \\
0.35 & -0.538850 & -0.538867 &  0.0383 & 0.0382 \\
0.4 & -0.523054 & -0.523259 &  0.0270 & 0.0268 \\
0.45 & -0.508895 & -0.508976 &  0.0173 & 0.0173 \\
0.5 & -0.496152 & -0.496289 &  0.0086 & 0.0086 \\
\end{tabular}
\end{ruledtabular}
\end{table}

Let us start discussing the ground-state energy, shown in Fig.~\ref{fig:energy}. For the unfrustrated case $J_2=0$, our results are fully 
compatible with what has been previously obtained in Refs.~\cite{rader2018,corboz2018}. The data points align perfectly according to the theoretical 
expectations and the extrapolated values are in very good agreement with quantum Monte Carlo results~\cite{sandvik1997,calandra1998}. For example 
for $D=7$ (after extrapolation in the environment dimension $\chi$) we get $e(D=7)=-0.669432$, which is identical to the linear extrapolation in 
$1/\xi^3$ from $D=3$ to $7$. Including the subleading term $1/\xi^4$, the extrapolation gives $e(\infty)= -0.669437(2)$ (to be compared with the 
exact value $e_{\rm QMC}=-0.669437(5)$~\cite{sandvik1997}). For future comparisons, the energies for $D=7$ and different $J_2/J_1$ ratios are 
reported in Table~\ref{tab:enemag}. Under increasing the frustrating ratio, a remarkably smooth behavior persists up to $J_2/J_1 \approx 0.3$; 
then, for larger values, small fluctuations on the fourth digit of the energy, are visible, possibily indicating that the scaling regime moves 
to larger values of $\xi$ (or $D$), not reachable within our current possibilities. Still, the quality of the results is sufficient to obtain 
reliable extrapolations for $\xi \to \infty$. Our calculations show that the expected scaling is not limited to the unfrustrated case, but 
persists in the whole antiferromagnetic region, thus corroborating the ideas put forward in Refs.~\cite{rader2018,corboz2018}. One remarkable 
feature is that, while for small values of $D$ (i.e., for $D=2$ and $3$), the correlation length $\xi$ clearly {\it increases} by increasing 
$J_2/J_1$, for larger values of $D$ (i.e., for $D=4$, $5$, $6$, and $7$), it is essentially constant, or even slightly decreasing with $J_2/J_1$. 
This aspect will be discussed in connection to the magnetization curve that is presented below.

\begin{figure}
\includegraphics[width=\columnwidth]{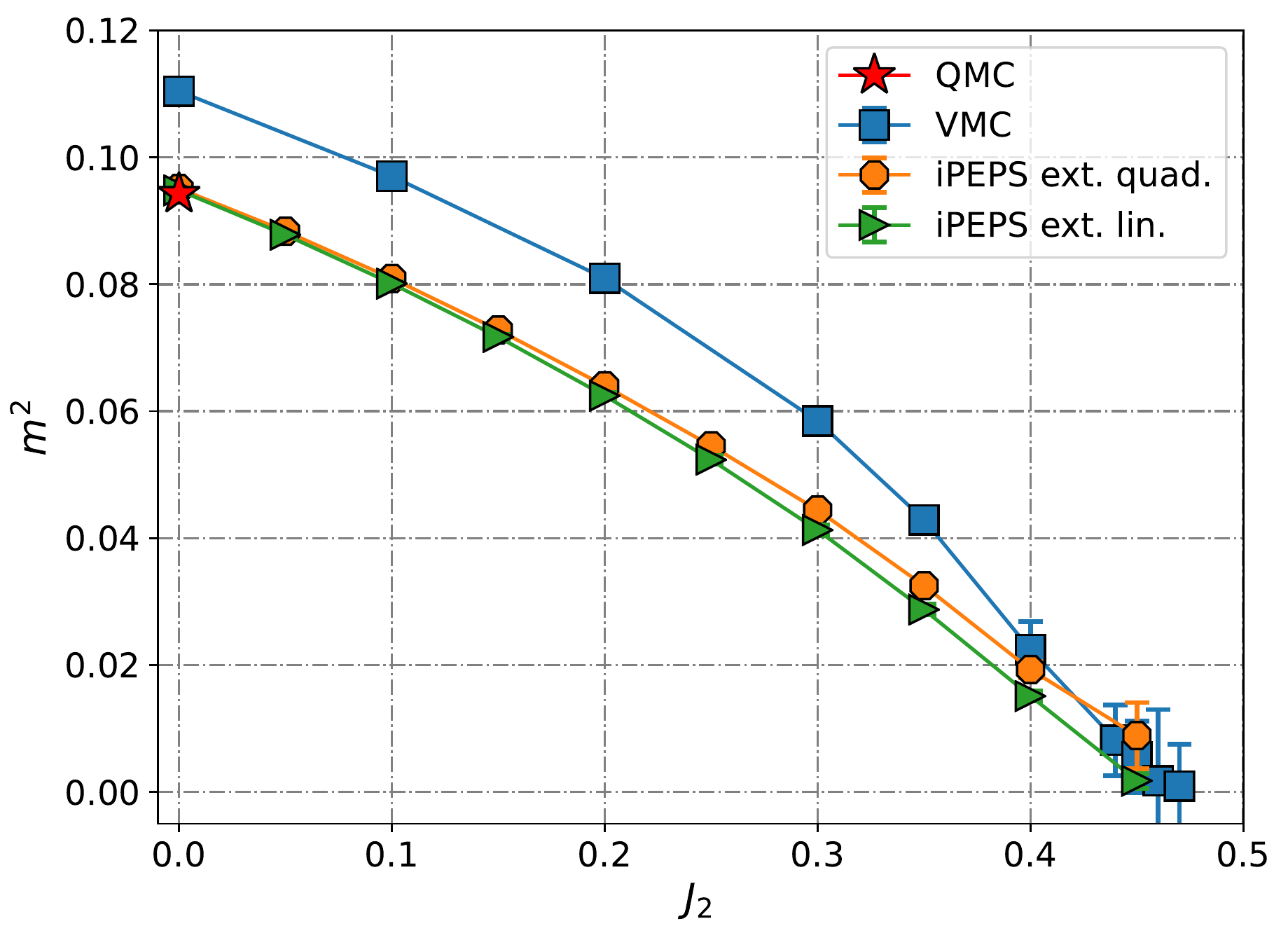}
\caption{\label{fig:final}
Magnetization (square) as a function of the frustrating ratio $J_2/J_1$ as obtained from Fig.~\ref{fig:magnet}. The exact result for $J_2=0$ 
is shown~\cite{sandvik1997}. For comparison, the variational Monte Carlo calculations of Ref.~\cite{ferrari2018} are also included.}
\end{figure}

\begin{figure}
\includegraphics[width=\columnwidth]{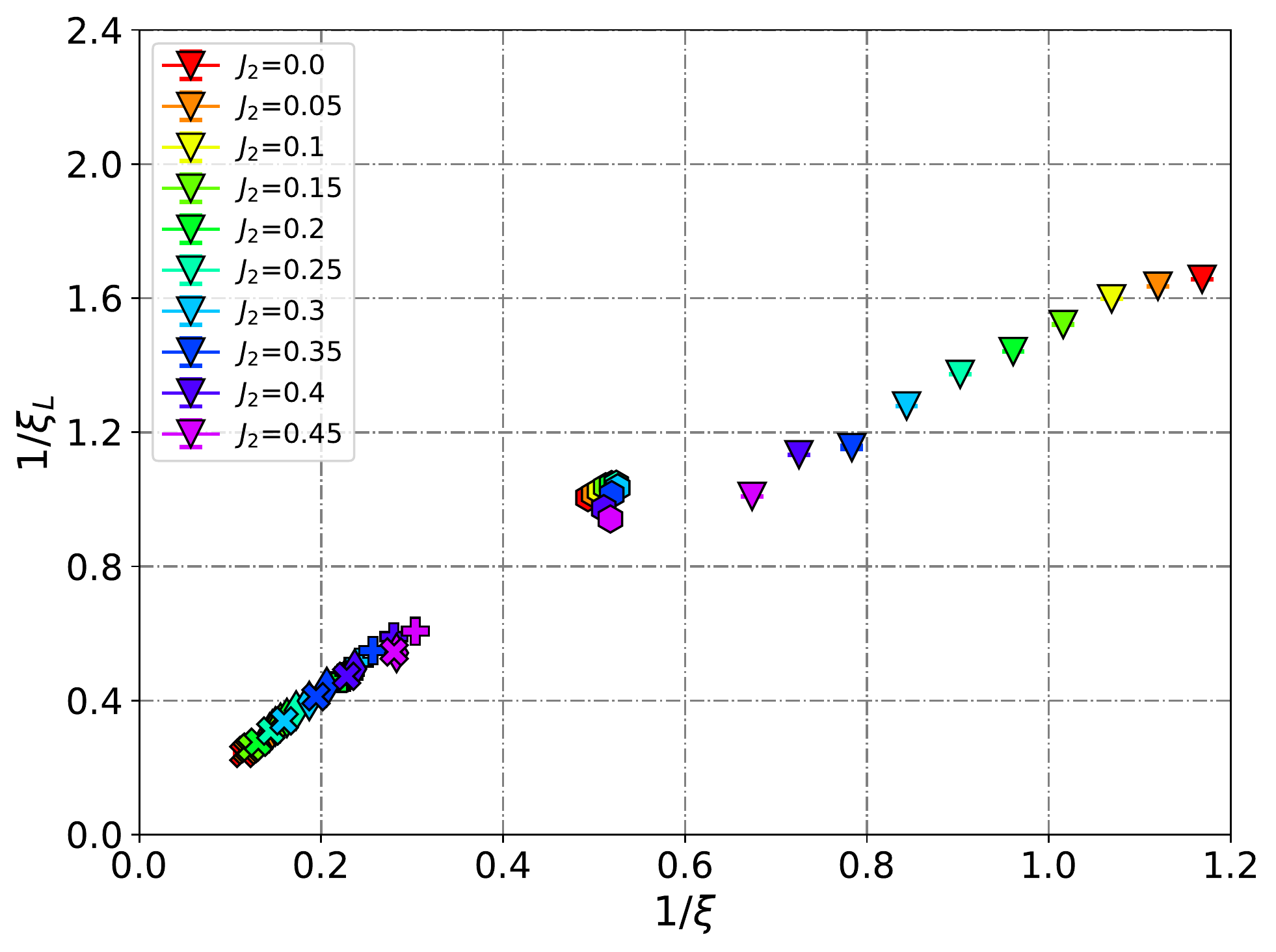}
\caption{\label{fig:xixi}
Longitudinal correlation length $\xi_L$, as extracted from the spin-spin correlations, as a function of the transverse one $\xi$ for different
values of $J_2/J_1$ at $D=3,\ldots,7$ (denoted by triangles, hexagons, pluses, diamonds, and crosses in the same order).}
\end{figure}

Then, we move to the central part of the present work, which deals with the magnetization, see Fig.~\ref{fig:magnet}. Here, we report $m^2(\xi)$ 
for different values of $J_2/J_1$ (including $0.5$) for $D$ ranging from $2$ to $7$. Furthermore, the raw data for $D=7$ are also shown in 
Table ~\ref{tab:enemag}. In the unfrustrated case, we get $m^2(D=7)=0.0994$ and $m^2(\infty)=0.0948(2)$, to be compared with the exact value 
$m^2_{\rm QMC}=0.0942(2)$~\cite{sandvik1997}. In Fig.~\ref{fig:magnet}, we attempt both linear and quadratic fits. As in the case of energy 
extrapolations, we exclude the results with $D=2$ from the fitting procedure, since they are clearly off, especially for intermediate and large 
values of $J_2/J_1$. According to our fits, the linear one looks more trustable than the quadratic one, which serves to give an upperbound to 
the value of the magnetization. Within the linear fit, we observe vanishing  magnetization for $J_2/J_1 \approx 0.46(1)$, giving rise to a
continuous transition to a magnetically disordered phase, whose nature is beyond the scope of the present work. We would like to emphasize that 
the results for $J_2/J_1=0.5$ are clearly incompatible with a smooth behavior in $1/\xi$, strongly suggesting that at this point the ground state 
is already outside the magnetically-ordered phase. The final magnetization curve is shown in Fig.~\ref{fig:final}. For comparison, the variational 
Monte Carlo calculations, which have been obtained by using Gutzwiller-projected fermionic states, are also shown~\cite{ferrari2018}. In the 
latter case, a quantum critical point for $J_2/J_1 \approx 0.48$, separating the antiferromagnetic phase and a gapless spin liquid, has been 
reported. The present results are expected to improve the accuracy of the magnetization (e.g., the accuracy of $m^2$ for the unfrustated case 
is smaller than $1\%$). Still, these two independent calculations give very similar behavior, with almost compatible values for the location 
of the quantum critical point. We would like to mention that, recent numerical calculations, including DMRG~\cite{wang2018}, neural-network 
approaches (based upon restricted Boltzmann machines on top of fermionic states)~\cite{nomura2020}, and finite size PEPS calculations~\cite{liu2020} 
also pointed out that the N\'eel phase survives up to $J_2/J_1$ in the range $0.45 \div 0.47$, a value that is considerably larger than the one 
predicted by linear spin-wave theory~\cite{chandra1988}. 

Finally, we would like to comment on the $J_2$-dependence of the correlation length, which is clearly different at small (i.e. $D=2,3$) and larger 
(i.e. $D=4,\cdots,7$) bond dimensions. A possible explanation of the rapid increase of $\xi$, for $D=2$ and $3$, when approaching the critical 
point, may be attributed to the fact that, for these very small bond dimensions, the antiferromagnetic state is poorly approximated as a ``dressed'' 
product state, having a finite magnetization but lacking the correct transverse (Goldstone) fluctuations. When approaching the phase transition, 
the magnetization decreases and the state starts to build up long-range entanglement (for $D=3$ a short-range resonating-valence bond state can be 
constructed~\cite{poilblanc2012}). Therefore, a larger correlation length can be attained. Once the basic (low $D$) structure of tensor is 
established, optimizing at increasingly higher $D$ further improves the description of the antiferromagnetic state and allows correlation length 
to grow, becoming large even in the presence of significant frustration. Then, no appreciable change of $\xi$ is detected when approaching the 
quantum critical point. In this respect, we expect that $\xi \to \infty$ in the whole N\'eel phase, including the critical point. Remarkably, 
despite optimized iPEPS being finitely correlated, the correct exponent of the power-law decay of transverse spin-spin correlations, i.e., 
$\langle S^x_0 S^x_r \rangle \simeq 1/r$ (assuming magnetization along $z$-spin axis), can already be obtained, see Appendix~\ref{appC} for 
the case with $J_2=0$.

As mentioned above, $\xi$ corresponds to the correlation length of transverse spin-spin correlations. In addition to that, it is possible to 
evaluate, by a direct fitting procedure of the correlation function itself, the correlation length $\xi_L$ of the longitudinal correlations. 
We find also this quantity to be relatively large, i.e., $\xi_L \approx \xi/2$, see Fig.~\ref{fig:xixi}. Moreover, as for transverse spin-spin 
correlations, the short-range behavior of the longitudinal correlations reveals their power-law decay (see Appendix~\ref{appC}), which then 
becomes rapidly cut off above the finite-$D$ induced length scale $\xi_L$. These findings show that our optimized iPEPS are even able to 
approximately capture the power-law behavior of transverse and longitudinal spin-spin correlations of the N\'eel phase. 

\section{Conclusions}\label{sec:concl}

In this work, we have investigated the antiferromagnetic phase of the spin-1/2 $J_1-J_2$ model on the square lattice, evaluating with unprecedent
accuracy the energies and magnetizations for $J_2/J_1 \le 0.45$. The results point towards the existence of a quantum critical point
at $J_2/J_1 \approx 0.46(1)$, which separate the N\'eel antiferromagnet and a quantum paramagnet, whose nature is beyond the scope of the present
study. The importance of our findings is twofold. From the methodological side, we combined state-of-the-art optimization techniques (based upon 
the AAD scheme~\cite{liao2019}), clever parametrizations of the tensor network (based upon the underlying residual $U(1)$ symmetry that exists in 
the N\'eel phase), and recently developed extrapolation analyses (based upon the correlation-length scaling~\cite{rader2018,corboz2018}).
In particular, the construction of $U(1)$-symmetric tensor is pivotal to a straight optimization procedure and correlation-length scaling to 
solid extrapolations to thermodynamic limit. With these tools in hand, it is possible to get reliable estimations for the ground-state energy but, 
most importantly, also for the magnetization within the frustrated regime, for which no exact methods can be applied. Therefore, the main outcome 
of the present work is to provide the magnetization curve for the spin-1/2 $J_1-J_2$ model on the square lattice up to relatively large values of 
the frustrating ratios. In particular, the magnetization curve shows a smooth behavior, which strongly suggest the existence of a continuous phase 
transition towards a quantum paramagnet.

Here, our calculations have been limited to the magnetically ordered phase, where relatively entangled states have been achieved. Indeed, rather
long correlation lengths are obtained, indication that the tensor network may approximately describe the existence of gapless excitations (i.e.,
Goldstone modes). The magnetically disordered phase still remains elusive, presumably because of its high-entangled nature due to fractional
excitations (spinons and visons). In this respect, the recently-developed method to impose $SU(2)$ symmetry~\cite{mambrini2016,poilblanc2017b} 
would be beneficial to the final understanding of the full phase diagram of the spin-1/2 $J_1-J_2$ model.

\begin{acknowledgments}
We thank Fabien Alet, Zhengcheng Gu, Andreas L{\"a}uchli, Wenyuan Liu, Pierre Pujol, Anders Sandvik, and Sandro Sorella for helpful discussions. 
This work was granted access to the HPC resources of CALMIP supercomputing center under the allocation 2020-P1231. 
This project is supported by the TNSTRONG ANR-16-CE30-0025 and the TNTOP ANR-18-CE30-0026-01 grants awarded by the French Research Council. 
\end{acknowledgments}

\appendix

\section{Smith normal form}\label{appA}

The Smith normal form of matrix $M$ is needed to solve the linear system introduced in Sec.~\ref{subsec:smith}. For a $n \times m$ integer matrix 
$M$ the Smith normal form is defined as 
\begin{equation}
LMR = S,
\end{equation}
with $L$ and $R$ being integer matrices with unit determinant and $n \times m$ integer matrix $S$. The only non-zero elements of $S$ are 
$S_{i,j} = s_i \delta_{i,i}$ for $1 \le i \le r$ where $r\le m$. These so-called {\it invariant factors} $s_i$ satisfy divisibility relations 
$s_i | s_{i+1}$ for $1 \le i < r$. The Smith Normal form conveniently reveals the vectors of integer charges $(\vec{u}$ and $\vec{v})$ spanning 
the $m-r$ dimensional kernel of the constraint system $M$ as the last $m-r$ columns of matrix $R$. Let us remark that such kernel vectors are 
unique up to an arbitrary multiples of trivial charge vectors $\vec{K}_0= [1,1,0,\dots,0]$ and $\vec{K}_1=[0,0,1,\ldots,1]$, as these merely 
move the constant $N$. In detail, a set of tensor elements $a^s_{uldr}$ satisfying $M \cdot (\vec{u},\vec{v}) = 0$ is identical to the set of 
elements satisfying $M \cdot [(\vec{u},\vec{v}) + \alpha \vec{K}_0 + \beta \vec{K}_1] = \alpha + 4\beta$ with $\alpha,\beta \in \mathds{Z}$.

\section{$1/D$ extrapolations}\label{appB}

\begin{figure}[t]
\includegraphics[width=\columnwidth]{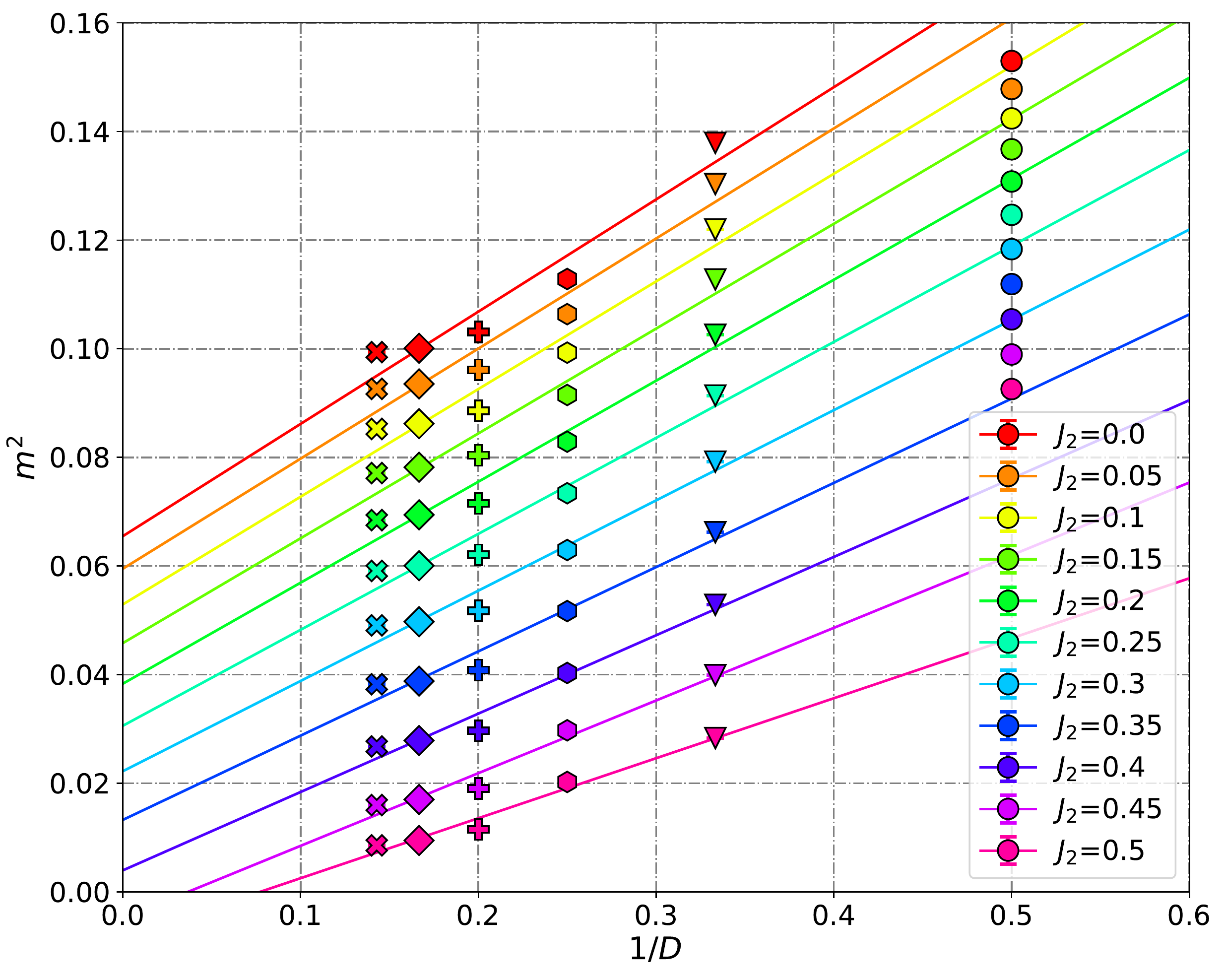}
\caption{\label{fig:magnet1D}
Linear extrapolation in $1/D$ for the $C_{4v}$-symmetric $U(1)$ iPEPS Ansatz with bond dimensions $D=2,\ldots,7$ denoted by circles, triangles, 
hexagons, pluses, diamonds, and crosses in the same order ($D=2$ data is excluded from the fit). Data is the same as in Fig.~\ref{fig:magnet}.}
\end{figure}

In Fig.~\ref{fig:magnet1D}, we report the results of the magnetization as a function of $1/D$. 
In this case, the magnetization cannot be described by a simple linear function in $1/D$
with appreciable accuracy. Considerable deviations are present, especially for larger bond dimensions across the studied range of $J_2/J_1$, 
preventing a smooth extrapolation in the $D \to \infty$ limit.

\section{Spin-spin correlations in the $J_2=0$ limit}\label{appC}

\begin{figure}[t]
\includegraphics[width=\columnwidth]{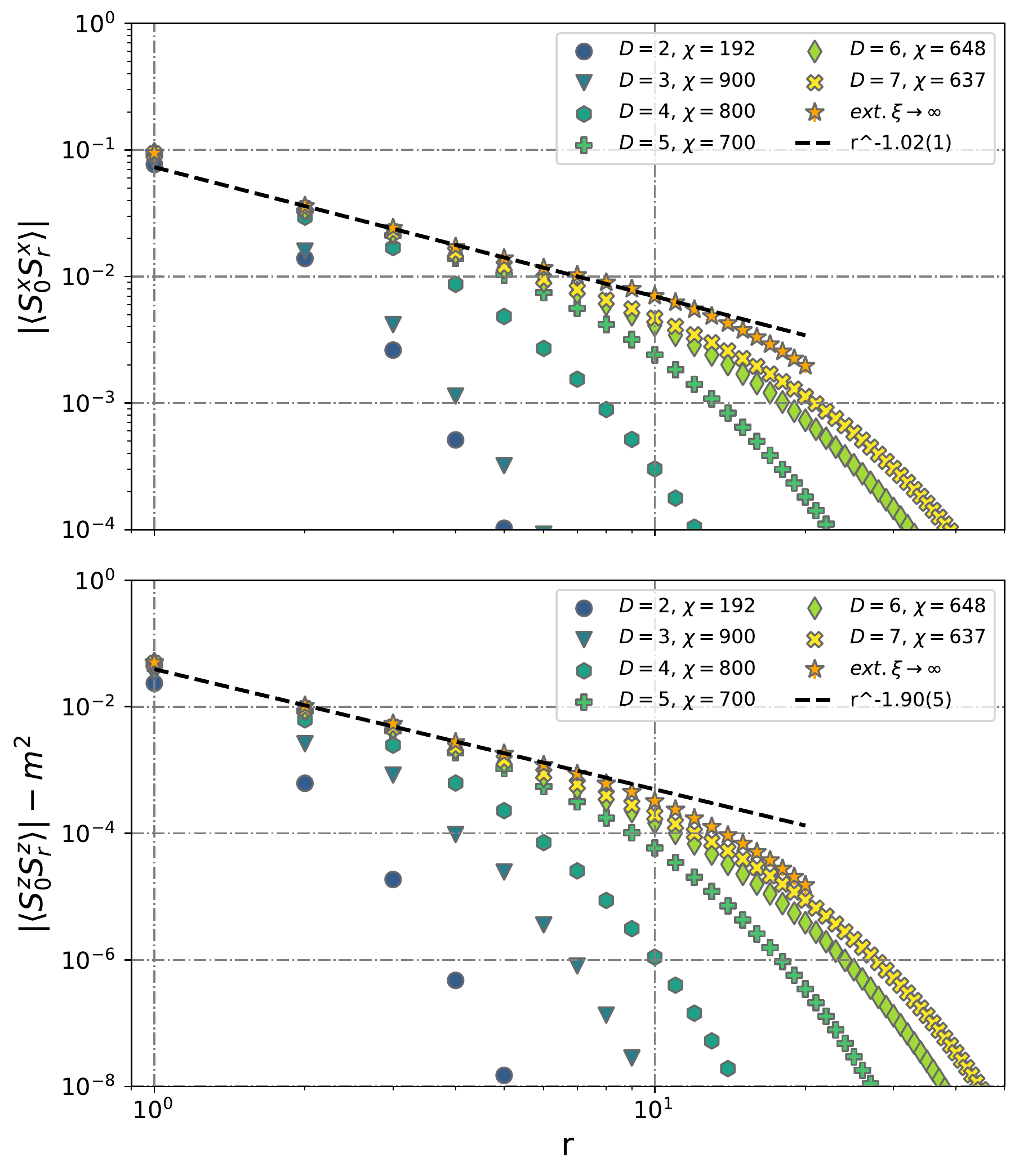}
\caption{\label{fig:xitj2=0}
Transverse (longitudinal) spin-spin correlations at $J_2=0$ are shown in the upper (lower) panel, for $D=2,\ldots,7$. Linear extrapolations in 
$1/\xi$, up to $r=20$, are performed using the $D=5,6,7$ data. The dashed lines are power-law fits to short-distance behavior, see text.}
\end{figure}

In Fig.~\ref{fig:xitj2=0}, assuming magnetization along $z$-spin axis, we show the decay of both transverse $\langle S^x_0 S^x_r \rangle$ 
and longitudinal $\langle S^z_0 S^z_r \rangle$ correlations for $J_2=0$ and $D=2,\ldots,7$. Due to imposed $U(1)$ symmetry the transverse 
correlations along $x$ and $y$ spin axes, $\langle S^x_0 S^x_r \rangle$ and $\langle S^y_0 S^y_r \rangle$, are identical. The extrapolated 
values are obtained by performing, for each distance $r$, an extrapolation in $1/\xi$ of the finite-$D$ results using the three largest 
available bond dimensions $D=5,6$, and $7$. Then, the extrapolated correlations are fitted in the short-distance region $r\in[2,11]$ (excluding 
the nearest-neighbor case) with a power law $f(r) \propto r^{-\beta}$. The final result gives an exponent $\beta \approx 1.02(1)$ for transverse 
and $\beta_L \approx 1.90(5)$ for longitudinal correlations.

\FloatBarrier

\nocite{*}
\bibliography{juraj1j2}

\end{document}